\documentclass[12pt,a4paper]{article}
\usepackage{amsmath,amsfonts}

\newtheorem{theorem}{Theorem}
\newtheorem{lemma}{Definition}

\begin{document}

\begin{titlepage}
\begin{center}

\textsc{\LARGE }\\[1.5cm]

\textsc{\Large }\\[0.5cm]

{ \Large \bfseries Analytical vectors and a new criterion of regularity for
representation of canonical commutation relations algebra\footnote{The XIXth International Workshop on High Energy Physics and Quantum Field Theory \\
8-15 September 2010 Golitsyno, Moscow, Russia.}}\\[0.4cm]

\textsc{\Large }\\[0.5cm]

\emph{Authors:}\\
\begin{itemize}
\item {M.\,N. Mnatsakanova\\
Skobeltsyn Institute of Nuclear Physics of Moscow State
University, Moscow, Russia.\\
E-mail: \texttt{mnatsak@theory.sinp.msu.ru}}

\item {S.\,G. Salynskiy\\
Department of Quantum Theory and High Energy Physics,
Faculty of Physics, Moscow State University, Moscow, Russia.}

\item {Yu.\,S. Vernov\\
Institute for Nuclear research of Russian Academy of Sciences, Moscow, Russia.\\
E-mail: \texttt{vernov@inr.ac.ru}}

\end{itemize}

\vfill

{\large June 2010}

\end{center}
\end{titlepage} 

\begin{abstract}

New criterion of regularity for representation of
canonical commutation relations algebra is given  on the  basis of
concept of an analytical vector.

\end{abstract}

\section{Introduction}

The algebra of canonical commutation relations (CCR) is a quantum
mechanics core, and its regular representations play a dominant
role in a science (various definitions of regularity for
representation of CCR algebras will be viewed further). The review
of CCR is given in a paper \cite{Putnam}. In the present article
we make a new definition of regularity for representation of CCR
algebras, using a concept of an analytical vector, introdused in
\cite{Nelson}.

In the most simple case of one dimension, CCR are defined as
follows:
\begin{equation}\label{1}
[\hat{p},\hat{q}] = -i\, \hat{I},
\end{equation}
where $\hat{p}$ and $\hat{q}$ are self-adjoint operators (in a
quantum mechanics they are impulse and coordinate operators accordingly).\\
As is known  in a case of finite number of operators, i.e. in a
case
$$
[\hat{p}_{i},\hat{q_{k}}] = - i\, \delta_{ik}; \qquad 1 \leq i,k
\leq n.
$$
all conclusions are similar to the results in a case of two
operators $\hat{p}$ and $\hat{q}$. Therefore we consider only the
case when equality (\ref{1}) is fulfilled. We  note that if there
is an infinite number of operators (the quantum field theory case),
the situation is more complicated and its viewing goes out for a
framework of the present article.

The Schrodinger representation is the most known of CCR
representations. It is realised in a space $L_2(-\infty,+\infty)$,
where functions $f(x)$ are such that
$\int^{+\infty}_{-\infty}|f(x)|^2dx<\infty$.  Operators $\hat{p}$
and $\hat{q}$ in the given representation are defined as follows:
\begin{equation}\label{2}
\hat{q}f(q) = qf(q), \qquad \hat{p}f(q) = -i\frac{d}{d\, q}f(q).
\end{equation}
The following definition of regularity for representation of CCR
algebras is widely used:
\begin{lemma}
Any representation of CCR algebras which is unitary equivalent to
Schrodinger one is regular.
\end{lemma}

There is one important circumstance: CCR algebras cannot be
realised by bounded operators  \cite{Putnam}, at least one of
operators $\hat{p}$ or $\hat{q}$ must be unbounded. We remind that
in the closed space unbounded operators are defined in a dense
domain.

In most papers CCR were investigated in Hilbert space. However, it
is possible to study CCR in spaces, which have an indefinite
metric \cite{MMSV}. Let's note that covariant gauge field theory
demands transition from a Hilbert space to a space with an
indefinite metric \cite{MS},\cite{KO}.

The Rellich-Dixmier's theorem is very important for the
description of the regular representations of CCR algebra. It
shows that representations of CCR algebra are regular for very
wide class of operators (see \cite{Putnam}).
\begin{theorem}
\emph{Rellich-Dixmier's theorem.}
Operators $\hat{q}$ and $\hat{p}$ form regular representation of CCR algebras if:\\
1. there exist dense domain $D\in D_q\bigcap D_p$ invariant under
the action
of $\hat{q}$ and $\hat{p}$ such that CCR hold on $D$;\\
2. the operator $(\hat{q}^2+\hat{p}^2)$ is essentially
self-adjoint on $D$.
\end{theorem}
Let's note that Fuglede has constructed an example of the
irregular representation when only requirement 1 is fulfilled
\cite{Fug}.

The representation of CCR algebras, realized by operators
(\ref{2}) in space $L_2(a,b)$, is an example of the irregular
representation.

It is possible to define CCR in the following form:
\begin{equation}\label{3}
[\hat{a},\hat{a}^*] = \hat{I},
\end{equation}
where operator $\hat{a}$ and adjoint operator $\hat{a}^*$ are
defined as
\begin{equation}\label{4}
\hat{a} = \frac{1}{\sqrt{2}}(\hat{q}-i\hat{p}) \quad \mbox{and}
\quad \hat{a}^* = \frac{1}{\sqrt{2}}(\hat{q}+i\hat{p}).
\end{equation}
It is easy to check that in a Hilbert space the spectrum of the
operator $\hat{N}=\hat{a}^*\hat{a}$ is $Sp\hat{N}=\mathbb{N}$ in
the regular representation.

The existence of a vacuum vector is a key feature of the regular
representations of CCR algebra in a Hilbert space. These
representations are known as Fock representations. It is obvious
that, cause of (\ref{1}) and (\ref{4}), a vector
$\psi_0=C\,e^{-\frac{q^2}{2}}$ satisfies  the requirement:
$\hat{a}\psi_0=0$, and hence $\hat{N}\psi_0=0$. Though all
definitions of regular representations are equivalent, some
representations are more convenient for research of CCR in spaces
which differ from a Hilbert one. The Krein space is an example of
such space which has an indefinite metric \cite{Bog},
\cite{Iokhv}. For example, the requirement of existence of an
eigenvector for the operator $\hat{N}$:
\begin{equation}\label{5}
\hat{N}\psi_{\alpha}=\alpha\psi_{\alpha}
\end{equation}
is one of definitions of regularity of representations in a Krein
space \cite{MMSV}.

In view of that operators $\hat{p}$ and $\hat{q}$, and naturally
$\hat{a}$ and $\hat{a}^*$, are unbounded there are some
difficulties related to definition of domains, in which they can
be determined. Use of the representation of CCR in a Weyl form
eliminates this difficulty:
\begin{equation}\label{6}
e^{it\hat{p}}e^{is\hat{q}}=e^{ist}e^{is\hat{q}}e^{it\hat{p}}
\end{equation}
It is well-known from the Stone's theorem that operators
$e^{it\hat{p}}$ and $e^{is\hat{q}}$ are bounded as operators
$\hat{p}$ and $\hat{q}$ are self-adjoint \cite{Ios}.

CCR in a Weyl form are widely used in a quantum mechanics (see,
for example, \cite{BR}). Until now CCR in this form were
considered in a Hilbert space, but it is natural to study a problem
of existence of a Weyl representation in a space with an
indefinite metric.

We hope what for these purposes, and probably more, a new
definition of regularity for representation of CCR algebra will be
very useful.

\section{Analytical vectors and their connection with CCR representation in a Weyl form}

Let's remember a definition of an analytical vector \cite{Nelson}.
\begin{lemma}
Let $\hat{A}$ be a linear operator on a Hilbert space $H$. A vector
$\xi\in H$ is called  analytic for $\hat{A}$ , if $\xi$ is in the
domain of $\hat{A^k}$ for every $k \in \mathbb{N}$ and for every
$t>0$
\begin{equation}\label{7}
\sum^{\infty}_{k=0}\frac{t^k}{k!}\|\hat{A}^k\xi\|<+\infty
\end{equation}
\end{lemma}

In this case we can define the operator $\exp{(t\,\hat{A})}$ as
its Taylor series
\begin{equation}\label{8}
e^{t\hat{A}}\xi = \sum^{\infty}_{k=0}\frac{t^k}{k!}\hat{A}^k\xi
\end{equation}
for all $\xi$ at which our series (\ref{8}) converges.

So, now we can formulate the main theorem.
\begin{theorem}
 Let's prove that a representation of CCR
algebras is regular, if there is a dense domain $D$, in witch any
vector $\xi\in D$ obeys conditions
\begin{equation}\label{9}
\sum^{\infty}_{k=0}\frac{t^k}{k!}\|\hat{q}^k\xi\|<\infty, \quad
\forall \, t>0; \quad
\sum^{\infty}_{k=0}\frac{s^k}{k!}\|\hat{p}^k\xi\|<\infty, \quad
\forall \, s>0
\end{equation}
and also any regular representation obeys conditions  (\ref{9}).
\end{theorem}

At first we will prove what this representation is regular. Our
proof is similar to the proof in \cite{Putnam}, but without the
assumption of  boundedness of operators $\hat{p}$ and $\hat{q}$.
From the relation (\ref{1}) it immediately follows that
\begin{equation}\label{10}
\hat{p}\,\hat{q}^n-\hat{q}^n\,\hat{p}=-i(\hat{q}^n)', \quad (' =
d/dq).
\end{equation}
Hence, with (\ref{10}) and (\ref{9}),
\begin{equation}\label{11}
\hat{p}\,e^{is\hat{q}}-e^{it\hat{p}}\,\hat{p}=-i(e^{it\hat{q}})'.
\end{equation}
From (\ref{11}) it directly follows
\begin{equation}\nonumber
e^{-it\hat{q}}\,\hat{p}\,e^{it\hat{q}}=(\hat{p}+t\hat{I})
\end{equation}
and,
\begin{equation}\label{12}
e^{-it\hat{q}}\,\hat{p}^n\,e^{it\hat{q}}=(\hat{p}+t\hat{I})^n.
\end{equation}
With condition (\ref{9}) we have what
\begin{equation}\label{13}
U_tV_s=e^{its}V_sU_t, \qquad U_t\equiv e^{it\hat{p}}, \; V_s\equiv
e^{is\hat{q}}.
\end{equation}
Thereby we have proved an existence of a Weyl relation (\ref{13})
in domain $D$.

The next step is a Weyl representations extension on a full space
$H$. For this purpose it is enough to note that $D$ is a dense
domain, and $U_t$ and $V_s$ are bounded operators that follows
from the Stone's theorem. We will note that the Stone's theorem
make some requirements on groups $U_t$ and $V_s$, but they are
weak (\cite{RSz-N}).

Now we will prove that if the relation (\ref{13}) is fulfilled in
$H$ then our representation is regular. But that part of task is
made for us by the von Neumann's theorem (see \cite{Putnam}). We
will note, as any regular representation is the direct sum of
irreducible representations, it is enough to view only an
irreducible representation.

Now we will show that any regular representation contains
analytical vectors in dense domain and, hence, satisfies  a Weyl
relation (\ref{13}) in $H$. For this purpose it is convenient to
use a CCR relation in the form (\ref{3}). In this case:
\begin{equation}\label{14}
\hat{q} = \frac{1}{\sqrt{2}}(\hat{a}+\hat{a}^*), \qquad \hat{p} =
\frac{1}{i\sqrt{2}}(\hat{a}-\hat{a}^*).
\end{equation}
Let us show that the operator $\hat{q}$, for example, has an
analytical vector in dense domain $D$. In the regular
representation we can construct an orthogonal basis, which
consists of eigenvectors $\psi_n$ of the operator
$\hat{N}=\hat{a}^*\hat{a}$. It is obviously from (\ref{3}) and
(\ref{5}) that $Sp\hat{N} = \mathbb{N}$. It is easy to show that:
\begin{equation}\label{15}
(\psi_n,\psi_n)=n!, \; \mbox{where} \; \psi_n =
{(\hat{a}^*)}^{n}\,\psi_0, \; \psi_0 \; \mbox{is a vacuum vector},
\; (\psi_0, \psi_0) = 1.
\end{equation}
and
\begin{equation}\label{16}
(\hat{a}^*\,\psi_{n},\, \hat{a}^*\,\psi_{n}) = {(n + 1)}!, \quad
(\hat{a}\,\psi_{n},\, \hat{a}\,\psi_{n}) = {n}! .
\end{equation}
Let us view a domain $D$, which consists of all finite linear
combinations of vectors $\psi_n$. As $H$ consists of all finite or
converging linear combinations of vectors $\psi_n$, then $D$ is a
dense domain. The norm of a vector $\psi_n$ can be set by the
formula $\|\psi_n\|=\sqrt{(\psi_n,\psi_n)}$. According to
(\ref{15})
\begin{equation}\label{17}
\|\psi_n\|=\sqrt{n!}.
\end{equation}
Accordingly, for a vector
\begin{equation}\label{18}
\psi=\sum^{m+n}_mC_k\, \psi_k
\end{equation}
from (\ref{15}) - (\ref{17}) we have the following restriction:
\begin{equation}\label{19}
\begin{gathered}
\|\hat{q}\psi\| \leq \sum^{m+n}_m\,\|C_k\,\hat{q}\,\psi_k\| \leq
\sqrt{2} C\,\sqrt{(m + n + 1)!}, \\ C = \max{\mid C_{k}\mid},
\quad m \leq k \leq m + n.
\end{gathered}
\end{equation}
For obtaining (\ref{19}) we have used that
\begin{gather*}
(\hat{a}^*\psi_n,\hat{a}^*\psi_n)=(\psi_n,(n+1)\,\psi_n)=(n+1)(\psi_n,\psi_n),\\
(\hat{a}\psi_n,\hat{a}\psi_n)=(\psi_n,n\psi_n)=n(\psi_n,\psi_n).\\
\end{gather*}
With (\ref{19}) at the end we have
\begin{gather}\label{20}
\sum^{\infty}_{k=0}\frac{t^k}{k!}\|\hat{q}^k\psi\|\leq
\sum^{\infty}_{k=0}\frac{t^k}{k!}\,C^{k}\,{(2\,(m + n +
1)!)}^{k/2}.
\end{gather}
It is obvious that the series (\ref{20}) converges for any finite
$m$ and $n$.

Thus, it is proved that the vector $\psi,\forall \psi\in D$ is
analytical for the operator $\hat{q}$. The proof of that fact that
any vector $\psi\in D$ is analytical for the operator $\hat{p}$
can be made by a similar way.

\section{Conclusion}

In the present article a new criterion of regularity for
representation of canonical commutation relations algebras is
given on the basis of concept of an analytical vector. We hope
that new definition will be useful for study a Weyl representation
in the indefinite metrics space.

\end{document}